\documentclass[12pt]{article}
\usepackage[T1]{fontenc}
\usepackage{graphicx}
\newcommand{\interlinia}{}

\title{Flexoelectric effect modeling}

\author{A. Kapanowski \\
{\em Institute of Physics, Jagiellonian University,}\\
{\em ulica Reymonta 4, 30-059 Cracow, Poland}  }

\begin{document}
\maketitle

\begin{abstract}
\interlinia
The statistical theory of the dipole flexoelectric (FE) polarization
in liquid crystals is used to calculate the temperature dependence
of order parameters, the elastic constants and the FE coefficients.
Two systems with polar wedge-shaped and banana-shaped molecules are
investigated.
In both cases the FE coefficients are proportional to the dipole
moment component parallel to the molecule symmetry axis.
The origin of the FE effect and
microscopic pictures of the distorted phases are discussed.
\newline\newline
{\bf Keywords:} nematic liquid crystals, flexoelectric polarization,
elastic constants, statistical theory.
\end{abstract}

\interlinia

\section{Introduction}

In an ideal nematic liquid crystal the molecules are on average aligned 
along one common direction $\pm\vec{N}$
\cite{[1993_de_Gennes]}.
There is usually some deformation of the alignment and the vector $\vec{N}$
depends on the position $\vec{r}$ in the liquid crystal.
The free-energy density due to the distortion of the vector $\vec{N}$
is expressed in terms of the vector derivatives and the elastic
constants $K_i$
\cite{[1994_Stallinga_Vertogen]}
\begin{equation}
\label{fd}
f_{d} =
{\frac {1}{2}} K_{1} (\nabla \cdot \vec{N} )^{2}
 + {\frac {1}{2}} K_{2} 
[ \vec{N} \cdot (\nabla \times \vec{N} ) ]^{2}
+ {\frac {1}{2}} K_{3} 
[ \vec{N} \times (\nabla \times \vec{N} ) ]^{2},
\end{equation}
where $K_{1}$, $K_{2}$, and $K_{3}$ are splay, twist, 
and bend elastic constants, respectively.

In a deformed uniaxial nematic liquid crystal, there should appear
in many cases a spontaneous dielectric polarization
invented by Meyer \cite{[1969_Meyer]}
\begin{eqnarray}
\vec{P} &=& e_1 \vec{N} (\nabla \cdot \vec{N})
+ e_3 (\vec{N} \cdot \nabla) \vec{N}
\nonumber\\
&=& \mbox{} e_1 \vec{N} (\nabla \cdot \vec{N})
- e_3 \vec{N} \times (\nabla \times \vec{N} ),
\end{eqnarray}
where $e_1$ and $e_3$ are the splay and the bend flexoelectric (FE)
coefficients, respectively.
The appearance of the spontaneous polarization in liquid crystals
as a result of orientational deformations is called
the flexoelectric effect.

The FE coefficients and the elastic constants are macroscopic
parameters that are important in practical applications of liquid crystals.
Thus, it is interesting to study the connection between microscopic
and macroscopic parameters in order to fasilitate new material production.
In this paper we will focus on the FE effect.

A microscopic mechanism of the FE effect was proposed by Meyer
\cite{[1969_Meyer]},
who showed that it is a steric effect due to the asymmetry of
the molecular shape. But Prost and Marcerou 
\cite{[1977_Prost_Marcerou]}
noticed that the polarization in a deformed liquid crystal 
can be also produced as a result of a gradient in the average density 
of quadrupole molecule moments.
In 1976 Straley developed a microscopic theory 
\cite{[1976_Straley]} by extending the Onsager
theory of orientational ordering in hard-rod gases.
Other mean-field theory was given by Derzhanski and Petrov
\cite{[1971_Derzhanski_Petrov]}.
The density-functional theory was derived by Singh and Singh 
\cite{[1989_Singh_Singh]},
were the dipole and the quadrupole contributions to the FE coefficients
were taken into account. Somoza and Tarazona
\cite{[1991_Somoza_Tarazona]} 
tried to include the effects of the relaxation
in the angular distribution function of the deformed nematic.
As far as computer simulations are concerned, we would like to mention
the papers by Stelzer {\em et al.}
\cite{[1999_Stelzer_Berardini_Zannoni]}
and by Billeter and Pelcovits
\cite{[2000_Billeter_Pelcovits]},
where pear-shaped molecules were studied.

In this paper, we present two systems with wedge-shaped
and banana-shaped molecules.
In our calculations, we use a simple potential energy of interactions
which allows us to obtain the macroscpic parameters of the systems
and to understand the FE effect origin in the deformed phases.

\section{Expressions for flexoelectric coefficients}

The microscopic free energy for the system can be derived 
in the thermodynamic limit 
($N\rightarrow\infty$, $V\rightarrow\infty$, $N/V=\mbox{const}$) 
from the Born-Bogoliubov-Green-Kirkwood-Yvon (BBGKY) hierarchy 
\cite{[1980_Reichl]}
or as the cluster expansion for the uniaxial systems
\cite{[1979_Stecki_Kloczkowski]}.
The total free energy $F$ consists of the entropy term
and the interaction term, namely
\begin{equation}
\label{Ftotal}
F = F_{ent} + F_{int},
\end{equation}
where
\begin{eqnarray}
\beta F_{ent} &=& 
\int {d\vec{r}}{dR} G(\vec{r},R)
 \{ \ln [ G(\vec{r},R) \Lambda ]-1 \},
\\
\beta F_{int} &=& 
- {\frac {1}{2}} \int {d\vec{r}_1}{dR_1}{d\vec{r}_2}{dR_2} 
G(\vec{r}_1,R_1) G(\vec{r}_2,R_2) f_{12}.
\end{eqnarray}
Here 
$f_{12}=\exp (-\beta \Phi_{12})-1 $  is the Mayer function, 
$\Phi_{12}$ is the potential energy of interactions,
$dR = d\phi d\theta \sin\theta d\psi$, 
$\beta = 1/(k_{B} T)$,
and $\Lambda$ is related to the ideal gas properties.
The normalization of the one-particle distribution function $G$ is
\begin{equation}
\label{normaG}
\int {d\vec{r}}{dR} G(\vec{r},R)=N. 
\end{equation}
The equilibrium distribution $G$ minimizing the free energy 
of Eq. (\ref{Ftotal}) satisfies the equation
\begin{equation}
\label{lnGgeneral}
\ln [G(\vec{r}_1,R_1) \Lambda]
- \int {d\vec{r}_2}{dR_2} G(\vec{r}_2,R_2) f_{12} = \mbox{const}.
\end{equation}
In the case of the homogeneous uniaxial nematic phase,
the distribution function depends only on the molecule orientation
$G(\vec{r},R) = G_0(R)$. We will further assume that $G$
depends on a single argument $G_0(R) = G_0(\vec{n} \cdot \vec{N})$, 
where $\vec{N}$ determines the phase orientation.
This is satisfied exactly for the molecules with
$C_{\infty v}$ symmetry but we expect that it is plausible
for the banana-shaped molecules with $C_{2v}$ symmetry.
The molecule electric dipole moment is defined as
\begin{equation}
\vec{\mu} = \mu_1 \vec{l} + \mu_2 \vec{m} + \mu_3 \vec{n},
\end{equation}
where the unit vectors $(\vec{l},\vec{m},\vec{n})$
describe the molecule orientation.

In the case of the wedge-shaped molecules, 
the FE coefficients have the form
\cite{[2007_Kapanowski]}
\begin{eqnarray}
e_1 &=& \mu_3
\int {d\vec{u}}{dR_1}{dR_2} f_{12} 
G_0(\vec{n}_1 \cdot \vec{N}) G_0'(\vec{n}_2 \cdot \vec{N})
u_x n_{1z} n_{2x},
\\
e_3 &=& \mu_3
\int {d\vec{u}}{dR_1}{dR_2} f_{12} 
G_0(\vec{n}_1 \cdot \vec{N}) G_0'(\vec{n}_2 \cdot \vec{N})
u_z n_{1x} n_{2x},
\end{eqnarray}
where $\vec{n}$ determines the molecule $C_{\infty v}$ symmetry axis.
In the case of the banana-shaped molecules,
the FE coefficients can be written in the approximated form 
provided that $\vec{l}$ determines the molecule $C_{2v}$ symmetry axis
and $\vec{n}$ determines the long molecule axis
\cite{[2008_Kapanowski]}
\begin{eqnarray}
e_1 &=& \mu_1
\int {d\vec{u}}{dR_1}{dR_2} f_{12} 
G_0(\vec{n}_1 \cdot \vec{N}) G_0'(\vec{n}_2 \cdot \vec{N})
u_x l_{1z} n_{2x},
\\
e_3 &=& \mu_1
\int {d\vec{u}}{dR_1}{dR_2} f_{12} 
G_0(\vec{n}_1 \cdot \vec{N}) G_0'(\vec{n}_2 \cdot \vec{N})
u_z l_{1x} n_{2x}.
\end{eqnarray}

In both cases, the FE coefficients are proportional to the dipole 
moment component parallel to the molecule symmetry axis.
It results from the symmetries of the interactions and 
of the Mayer function.
The elastic constants of the uniaxial nematic phase
are \cite{[1997_Kapanowski]}
\begin{eqnarray}
\beta K_1 = \frac{1}{2} \int {d\vec{u}} {dR_1}{dR_2}
f_{12} G_0'(\vec{n}_1 \cdot \vec{N}) G_0'(\vec{n}_2 \cdot \vec{N}) 
u_x^2 n_{1x} n_{2x},
\\
\beta K_2 = \frac{1}{2} \int {d\vec{u}} {dR_1}{dR_2}
f_{12} G_0'(\vec{n}_1 \cdot \vec{N}) G_0'(\vec{n}_2 \cdot \vec{N}) 
u_y^2 n_{1x} n_{2x},
\\
\beta K_3 = \frac{1}{2} \int {d\vec{u}} {dR_1}{dR_2}
f_{12} G_0'(\vec{n}_1 \cdot \vec{N}) G_0'(\vec{n}_2 \cdot \vec{N}) 
u_z^2 n_{1x} n_{2x}.
\end{eqnarray}

\section{Results}

The calculations were performed for the square-well potential
energy of the form
\begin{equation}
\Phi_{12}(u/ \sigma) =  
\left\{
\begin{array}{lll}
+\infty   & \mbox{for} & (u/\sigma) < 1,  \\
-\epsilon & \mbox{for} & 1 < (u/\sigma) < 2,  \\
0         & \mbox{for} & (u/\sigma) > 2,
\end{array}  
\right.
\end{equation}
where  $u$ is the distance between molecules,
$\vec{r}_2 - \vec{r}_1 = u\vec{\Delta}$,
and $\sigma$ depends on the molecule orientations 
and on the vector $\vec{\Delta}$.
For $\sigma$, we used simple expressions with three parameters,
$\sigma_0$ defines the length scale,
$\sigma_1$ defines the nematic term, and
$\sigma_2$ defines the FE term.
In the case of the wedge-shaped molecules $\sigma$ has the form
\begin{equation}
\label{sigma-stozek}
\sigma = \sigma_{0} + \sigma_{1} 
\left[
(\vec{\Delta} \cdot \vec{n}_{1})^{2}+(\vec{\Delta} \cdot \vec{n}_{2})^{2}
\right] 
+ \sigma_{2}  (\vec{\Delta} \cdot \vec{n}_1 - \vec{\Delta} \cdot \vec{n}_2),
\end{equation}
whereas for the banana-shaped molecules 
\begin{equation}
\label{sigma-banan}
\sigma = \sigma_{0} + \sigma_{1} 
\left[
(\vec{\Delta} \cdot \vec{n}_{1})^{2}+(\vec{\Delta} \cdot \vec{n}_{2})^{2}
\right] 
+ \sigma_{2} (\vec{\Delta} \cdot \vec{l}_1 - \vec{\Delta} \cdot \vec{l}_2).
\end{equation}
We used the density $N V_{mol}/V = 0.1$,
$\sigma_1 = 2\sigma_0$, $\sigma_2 = \sigma_0$.
The molecule volume $V_{mol}$ was estimated from the mutually
excluded volume.
The FE coefficients were expressed in $\mu_i/\sigma_0^2$,
the elastic constants in $\epsilon/\sigma_0$, and
the temperature in $\epsilon/k_B$.
The potential energy of the form $\Phi_{12}(u/\sigma)$
allows us to express the FE coefficients and the elastic constants
as a finite series of the order parameters where the coefficients
of the expansion can be calculated analitically.
As a consequence it is easy to change the set of parameters
$\sigma_i$ and to study different systems.

The temperature dependence of the order parameters and the elastic 
constants is similar to the known results
\cite{[2007_Kapanowski]}, therefore the FE coefficients 
will be discussed only.

\subsection{Wedge-shaped molecules}

\begin{figure}
\begin{center}
\includegraphics{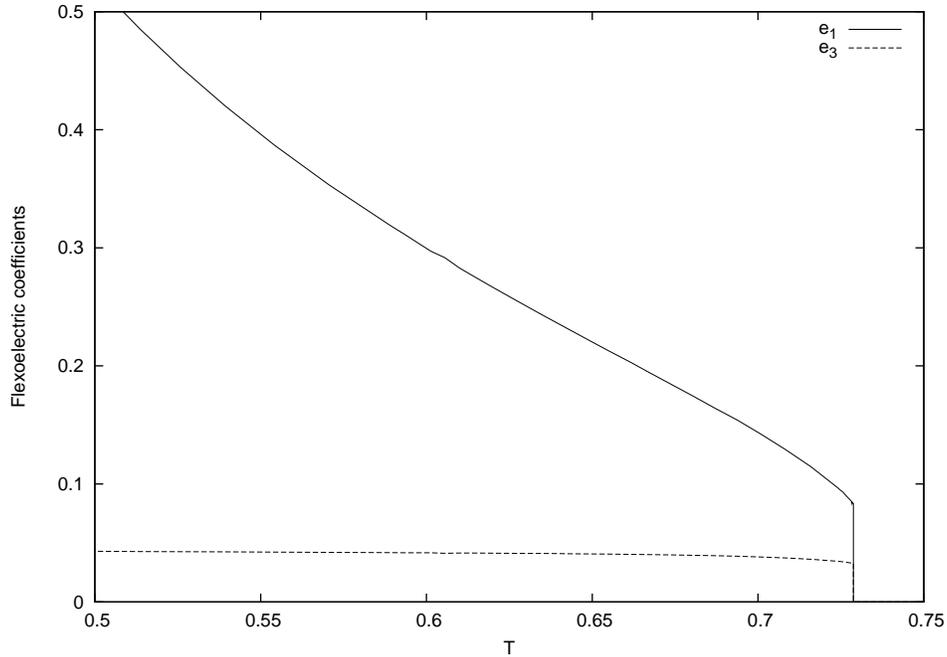}
\end{center}
\caption[The temperature dependence of the FE coefficients 
(wedge-shaped molecules).]{
\label{fig1}
\interlinia
The temperature dependence of the flexoelectric coefficients $e_i$
expressed in $\mu_3/\sigma_0^2$ in the case of wedge-shaped molecules. 
The temperature $T$ is expressed in $\epsilon/k_B$.
The $I-N$ transition is at $T_C=0.7287$.}
\end{figure}

\begin{figure}
\begin{center}
\includegraphics[scale=0.5]{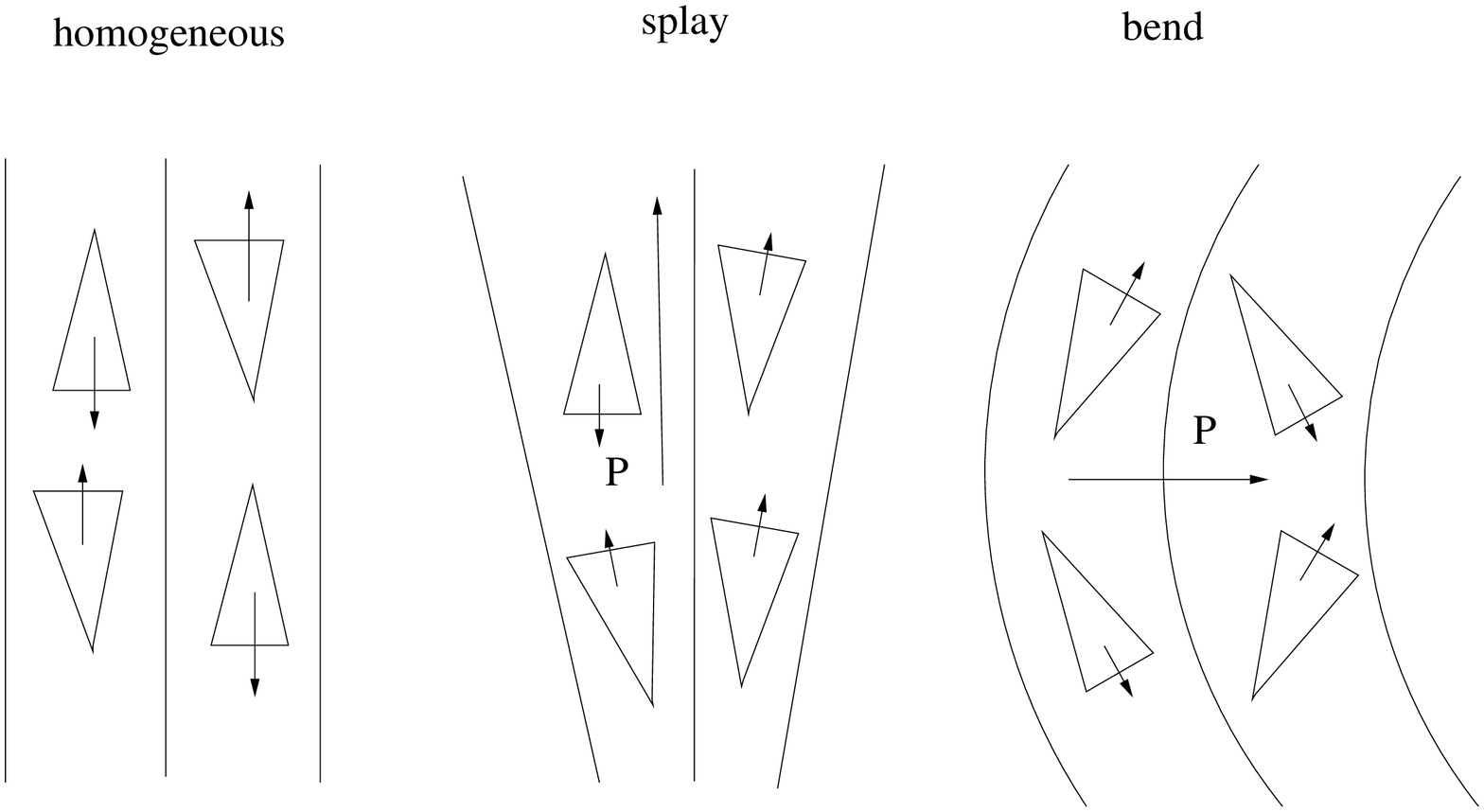}
\end{center}
\caption[The homogeneous, the splay, and the bend phases 
of wedge-shaped molecules.]{
\label{fig2}
\interlinia
The homogeneous, the splay, and the bend phases of wedge-shaped molecules.
The solid lines denote the local direction $\vec{N}$.
The $z$ axis is the symmetry axis of the homogeneous nematic phase.
In the splay phase there is an excess of molecules pointing down 
to the splay origin. 
In the bend phase there is an excess of molecules pointing to the left 
whereas the bend center is on the right.}
\end{figure}

In the system of the wedge-shaped molecules,
on decreasing the temperature we meet the first order
transition from the isotropic to the uniaxial nematic phase 
at $T_C= 0.7287$.
The temperature dependence of the FE coefficients
is presented in Fig.~\ref{fig1}.
The physical picture is shown in Fig.~\ref{fig2}.
On decreasing the temperature,
the bend coefficient $e_3$ is almost constant
while the splay coefficient $e_1$ increases monotonically.
The wedge-shaped molecules fit to the splayed structure
and there is an excess of molecules pointing to the splay origin.
The induced polarization has the opposite direction.
In the bend structure the FE polarization points to the bend centre
because the molecules are slightly turned in the opposite direction.

\subsection{Banana-shaped molecules}

\begin{figure}
\begin{center}
\includegraphics{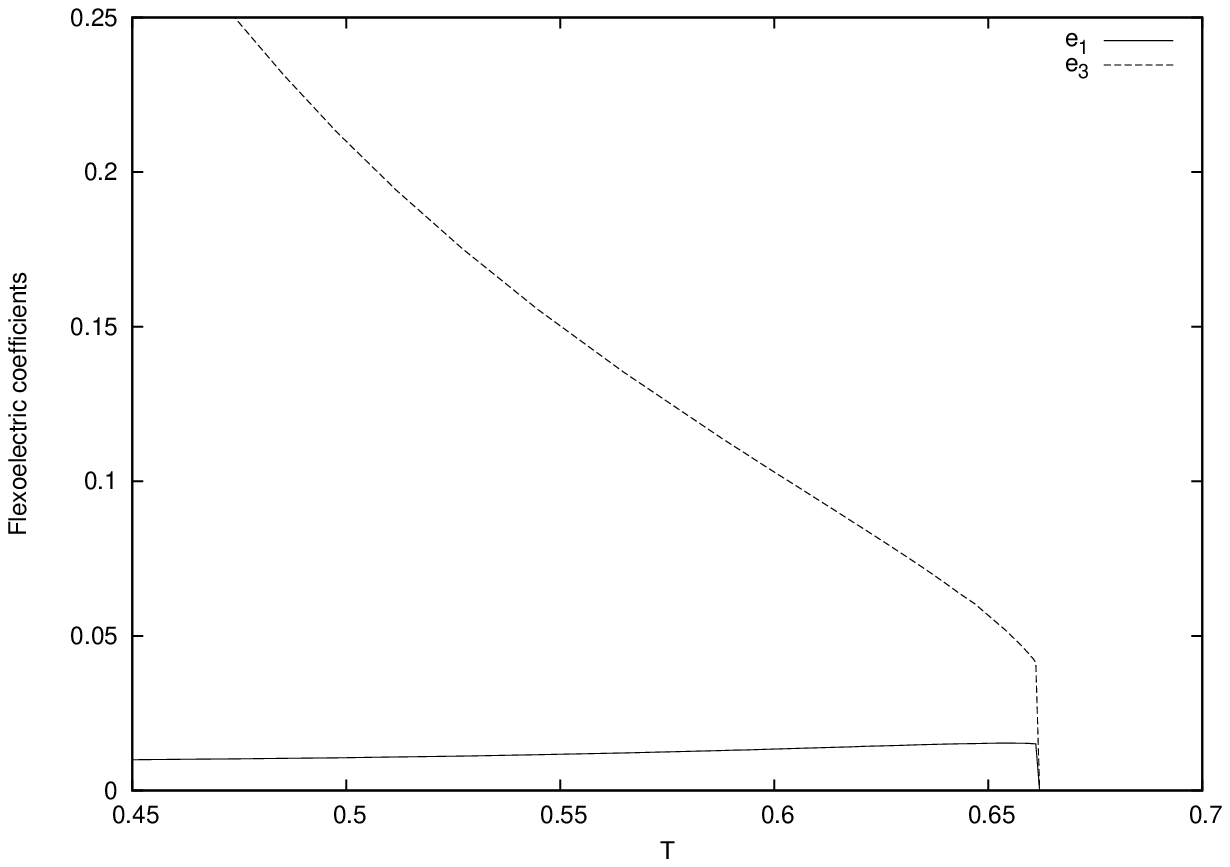}
\end{center}
\caption[The temperature dependence of the FE coefficients 
(banana-shaped molecules).]{
\label{fig3}
\interlinia
The temperature dependence of the flexoelectric coefficients $e_i$
expressed in $\mu_1/\sigma_0^2$ in the case of banana-shaped molecules. 
The temperature $T$ is expressed in $\epsilon/k_B$.
The $I-N$ transition is at $T_C=0.6611$.}
\end{figure}

\begin{figure}
\begin{center}
\includegraphics[scale=0.5]{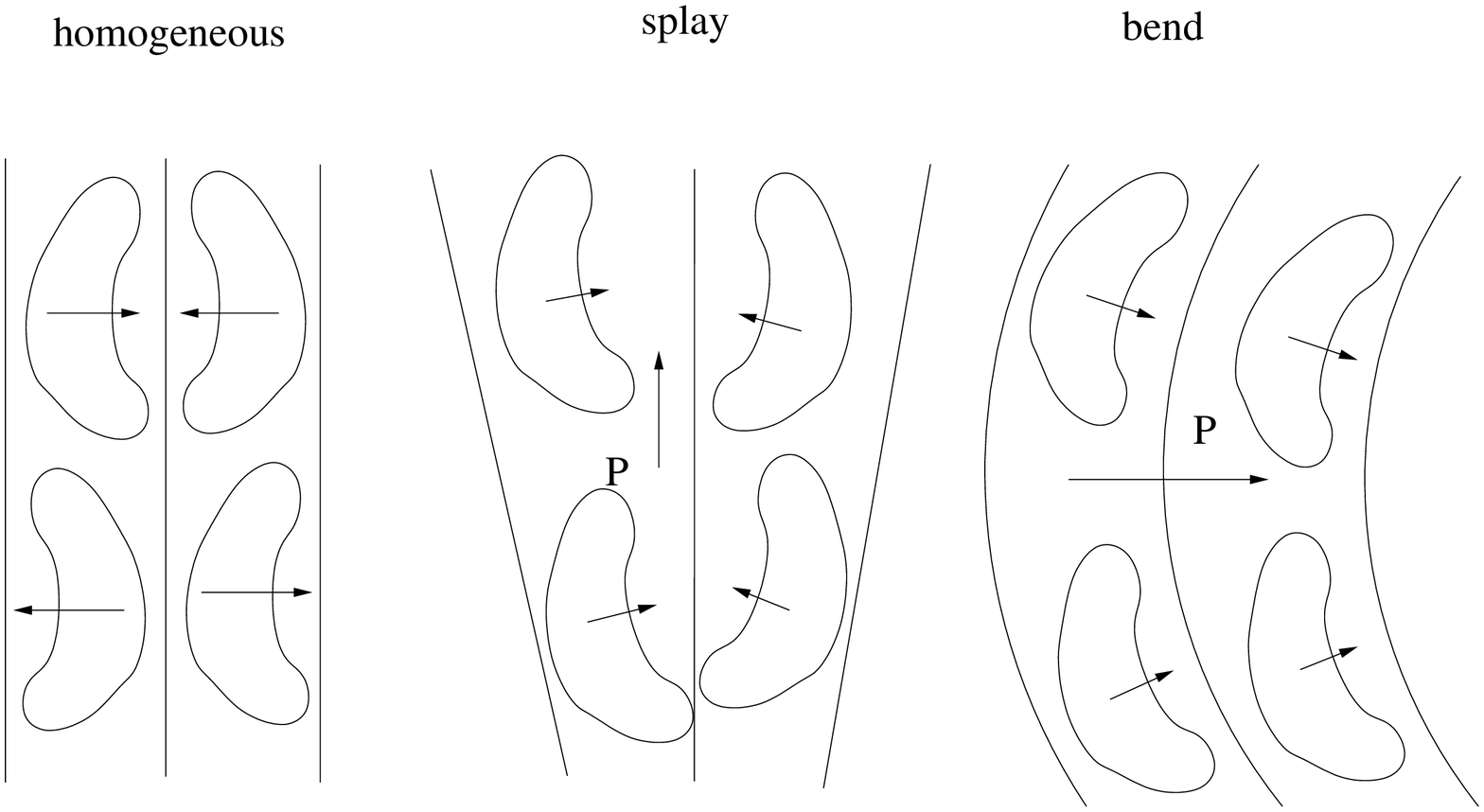}
\end{center}
\caption[The homogeneous, the splay, and the bend phases 
of banana-shaped molecules.]{
\label{fig4}
\interlinia
The homogeneous, the splay, and the bend phases of banana-shaped molecules.
The solid lines denote the local direction $\vec{N}$.
The $z$ axis is the symmetry axis of the homogeneous nematic phase.
In the splay phase, molecules are slightly turned and the average 
polarization direction is opposite to the splay origin. 
In the bend phase, the molecule dipole moments point 
to the bend center on the right.}
\end{figure}

In the system of the banana-shaped molecules,
on decreasing the temperature we meet the first order
transition from the isotropic to the uniaxial nematic phase 
at $T_C= 0.6611$.
The temperature dependence of the FE coefficients
is presented in Fig.~\ref{fig3}.
The physical picture is shown in Fig.~\ref{fig4}.
The banana-shaped molecules fit to the bend structure
and the FE polarization points to the bend centre.
In the splayed structure the molecules are slightly turned
and the average polarization is opposite to the splay centre position.

\section{Conclusions}

In this paper, the statistical theory was used to study the dependence
between the microscopic and macroscopic parameters of nematic liquid
crystals.
The temperature dependence of the order parameters, the FE coefficients,
and the elastic constants was obtained for the two systems.
For the wedge-shaped molecules
the bend coefficient $e_3$ is almost constant whereas
the splay coefficient $e_1$ is changing monotonically.
For the banana-shaped molecules
the splay coefficient $e_1$ is almost constant whereas
the bend coefficient $e_3$ is changing monotonically.
In both cases the FE coefficients are proportional to the molecule dipole
moment component, parallel to the molecule symmetry axis.
This is the long axis of the wedge-shaped molecules
and the short axis of the banana-shaped molecules.
It results from the symmetries of the interactions 
and of the Mayer function. In other theories, where the interactions
are not explicitly used or they are more complicated,
the dependence on the all dipole moment components is present.
The results are consistent with the microscopic pictures from the paper 
by Meyer but two additional situations were taken into account,
the splayed phase with banana-shaped molecules 
and the bend phase with wedge-shaped molecules.

At present stage, the qualitative comparison between the theory
and the experiment is very difficult.
The experimental data on the FE coefficients are still scarce
and sometimes contradictory
\cite{[2001_Petrov]}.
On the other hand, a nontrivial dependence on the details 
of the chemical structure was shown
\cite{[2001_Ferrarini]}.
A small chemical modification of the molecule can generate
significant change of the flexoelectricity.
Thus, we would like to give only the estimates assuming
typical values of the model parameters.
If we assume the molecular length $\sigma_0=1nm$,
the interaction energy $\epsilon-0.1 eV$,
and the electric dipole moment $\mu=1D$,
then we can estimate the values of the FE coefficients
$\mu/\sigma_0^2 = 3.3 pC/m$,
the elastic constants $\epsilon/\sigma_0 = 16 pN$,
and the temperature $\epsilon/k_B = 1160K$
(it is three/four times larger then the typical
isotropic-nematic transition temperature).

The theory should be improved in order to describe exactly real
systems. For the distance $\sigma$, the general expansion 
proposed by Blum and Torruella
\cite{[1972_Blum_Torruella]}
can be applied in order to describe molecular shapes better.
The Meyer function $f_{12}$ used in the low density
limit can be replaced with the direct correlation
function $c_{12}$ from the Percus-Yevick approximation.
In many cases, liquid crystals are mixtures of different
substances or different conformers and it should be taken
into account. Finally, it is possible to consider the FE effect 
in the biaxial nematic phase although this phase is still very rare.

\section*{Acknowledgements}

This work was supported by the Faculty of Physics, Astronomy and
Applied Computer Science, Jagellonian University (WRBW No. 45/06).
The author is also grateful to J. Spa{\l}ek for his support.


\end{document}